\documentclass[twocolumn]{aastex62}

\usepackage{graphicx}
\usepackage{xcolor}
\usepackage{xspace}
\usepackage{amsmath}
\usepackage[sort&compress]{natbib}
\usepackage[hang,flushmargin]{footmisc}

\usepackage{hyperref}
\definecolor{niceblue}{rgb}{0.0, 0.4, 0.65}
\definecolor{linkcolor}{RGB}{62, 101, 158}
\definecolor{citecolor}{rgb}{0.4,0.4,0.4}
\hypersetup{colorlinks=true,linkcolor=linkcolor,citecolor=citecolor,
            filecolor=linkcolor,urlcolor=linkcolor}
\hypersetup{pageanchor=false}

\newcommand{\acronym}[1]{{\small{#1}}}
\newcommand{\project}[1]{\textsl{#1}}
\newcommand{\code}[1]{{\texttt{#1}}}

\definecolor{nicered}{RGB}{255, 77, 77}
\newcommand{\changed}[1]{\textcolor{black}{#1}}

\newcommand{\Lone}[2]{\ensuremath{#1\| \mvec{#2} \|^1_1}}
\newcommand{\Ltwo}[2]{\ensuremath{#1\| \mvec{#2} \|^2_2}}

\newcommand{\mmatrix}[1]{\ensuremath{\boldsymbol{#1}}}
\newcommand{\mvec}[1]{\ensuremath{\boldsymbol{#1}}}

\newcommand{\teff}{\ensuremath{T_{\rm eff}}\xspace}
\newcommand{\logg}{\ensuremath{\log g}\xspace}

\newcommand{\mh}{\ensuremath{\mathrm{[M/H]}}\xspace}

\newcommand{\kms}{\ensuremath{\mathrm{km}~\mathrm{s}^{-1}}\xspace}
\newcommand{\ms}{\ensuremath{\mathrm{m}~\mathrm{s}^{-1}}\xspace}

\newcommand{\ang}{\text{\normalfont\AA}\xspace}
\newcommand{\SNR}{\acronym{S/N}\xspace}

\newcommand{\TF}{\code{TensorFlow}\xspace}
\newcommand{\python}{\code{python}\xspace}
\newcommand{\wobble}{\code{wobble}\xspace}
\newcommand{\starry}{\code{\acronym{starry}}\xspace}
\newcommand{\HARPS}{\project{\acronym{HARPS}}\xspace}

\newcommand{\RV}{\acronym{RV}\xspace}
\newcommand{\RVs}{\acronym{RV}s\xspace}
\newcommand{\EPRV}{\acronym{EPRV}\xspace}
\newcommand{\BERV}{\acronym{BERV}\xspace}
\newcommand{\PHOENIX}{\project{\acronym{PHOENIX}}\xspace}
\newcommand{\RM}{Rossiter-McLaughlin effect\xspace}
\newcommand{\LSF}{\acronym{LSF}\xspace}
\newcommand{\PCA}{\acronym{PCA}\xspace}

\newcommand{\gaia}{\emph{Gaia}\xspace}

\newcommand{\flatiron}{Flatiron Institute, Simons Foundation, 162 Fifth Ave, New York, NY 10010, USA}
\newcommand{\chicago}{Department of Astronomy and Astrophysics, University of
Chicago, 5640 S. Ellis Ave, Chicago, IL 60637, USA}
\usepackage{ bbold }
\newcommand{\Mdwarf}{Barnard's Star\xspace} 

\begin{document}\sloppy\sloppypar\raggedbottom\frenchspacing 

\shorttitle{\wobble}
\shortauthors{Bedell et al.}

\DeclareGraphicsExtensions{.pdf,.eps,.png}


\title{\textsc{\wobble: a data-driven \changed{analysis technique for time-series stellar spectra}}}

\author[0000-0001-9907-7742]{Megan Bedell}
\affiliation{\flatiron}

\author[0000-0003-2866-9403]{David W. Hogg}
\affiliation{\flatiron}
\affiliation{Center for Cosmology and Particle Physics, Department of Physics, New York University, 726 Broadway, New York, NY 10003, USA}
\affiliation{Center for Data Science, New York University, 60 Fifth Ave, New York, NY 10011, USA}
\affiliation{Max-Planck-Institut f\"ur Astronomie, K\"onigstuhl 17, D-69117 Heidelberg}

\author[0000-0002-9328-5652]{Daniel Foreman-Mackey}
\affiliation{\flatiron}

\author[0000-0001-7516-8308]{Benjamin T. Montet}
\altaffiliation{Sagan Fellow}
\affiliation{\chicago}

\author[0000-0002-0296-3826]{Rodrigo Luger}
\affiliation{\flatiron}

\correspondingauthor{Megan Bedell}
\email{E-mail: mbedell@flatironinstitute.org}

\begin{abstract}

\changed{In recent years, dedicated extreme-precision radial velocity (\EPRV) spectrographs have produced vast quantities of high-resolution, high-signal-to-noise time-series spectra for bright stars. 
These data contain valuable information for the dual purposes of planet detection via the measured \RVs and stellar characterization via the co-added spectra. 
However, considerable data analysis challenges exist in extracting these data products from the observed spectra at the highest possible precision, including the issue of poorly-characterized telluric absorption features and the common use of an assumed stellar spectral template. 
In both of these examples, precision-limiting reliance on external information can be sidestepped using the data directly. } 
Here we propose a data-driven method to simultaneously extract precise \RVs and infer the underlying stellar and telluric spectra using a linear model (in the log of flux). 
The model employs a convex objective and convex regularization to keep the optimization of the spectral components fast. 
We implement this method in \wobble, an open-source \python package which uses \TF in one of its first non-neural-network applications to astronomical data. 
In this work, we demonstrate the performance of \wobble on archival \HARPS spectra.
We recover the canonical exoplanet 51 Pegasi b, detect the secular \RV evolution of the M dwarf \Mdwarf, and retrieve the \RM for the Hot Jupiter HD 189733b. 
The method additionally produces extremely high-\SNR composite stellar spectra and detailed time-variable telluric spectra, which we also present here. 
\end{abstract}


\section{Introduction}

Precise radial velocity (\RV) measurements are critical to the discovery and characterization of exoplanets. 
On the order of one dozen dedicated spectrographs exist for the purpose of \RV planet-hunting, with at least as many more currently under construction \citep{Wright2017}. 
However, significant challenges exist in deriving precise \RV measurements from these spectra. 

One \changed{major contributor} to the noise budget in \RV measurements is the incomplete treatment of telluric features in the Earth's atmosphere \citep{Halverson2016}. 
Often, particular sections of a spectrum that are likely to feature telluric features are identified before the velocity shift of the stellar spectrum is inferred. 
These regions are then removed from analysis, leaving only seemingly telluric-free regions to be analyzed \citep[e.g.][]{AngladaEscude2012}.

This method has two significant issues. 
The first is that removing sections of the spectrum can remove significant amounts of information about the star, lowering the precision at which we can measure the stellar radial velocity. 
Many of the regions of significant telluric absorption lie in the red-optical and near-infrared, where there are abundant narrow spectral features that can be used to improve \RV precision \citep{Bottom2013}. 
This is especially true for M dwarfs, which peak in emitted energy at $\approx 1 \mu$m and have many narrow molecular absorption features in their photospheres \citep{Figueira2016}. 
Eliminating large chunks of these spectra will therefore significantly inhibit our ability to detect planets around M dwarfs through \RVs.

Secondly, not all telluric features are obvious. 
The Earth's atmosphere induces many small-amplitude features, often referred to as ``microtellurics,'' which are not obvious by eye but can affect the star's inferred \RV at the $\sim 1$ \ms level \citep{Cunha2014}. 
As the locations of these features are not known \textit{a priori} and may not even be apparent in stacked spectra of many observations, these spectral regions cannot be thrown out. 
Instead, alternative methods to account for these features as a part of the model must be developed and employed in order to mitigate the effect of the Earth's atmosphere on the measured stellar radial velocities.

One such approach is modeling the telluric spectrum using existing line databases like \acronym{HITRAN} \citep{HITRAN2016}. 
The telluric model may then be divided out from the observations, assuming the line spread function of the instrument is known \citep[e.g.][]{Seifahrt2010}. 
This method relies on existing physical knowledge about the Earth's atmosphere and can be fine-tuned using local observatory measurements of e.g. atmospheric water vapor content \citep{Baker2017}. 
However, line databases are incomplete even in significant absorption features when compared to actual observations and certainly do not include microtellurics, making them poorly suited for extreme precision RV applications \citep{Bertaux2014}.

Another option is the use of telluric standard observations: a spectrum of a rapidly rotating early-type star, which is virtually featureless due to extreme rotational line broadening, may be used as a telluric model and divided out. 
This approach has the advantage of naturally reproducing the instrumental line profile and current observing conditions if the standard star has a line-of-sight vector sufficiently close to the target and if both observations are taken close together in time. 
For these conditions to be true, though, requires a significant investment of observing time, which planet search programs often cannot afford. 
Additionally, artifacts may remain near strong telluric features due to the imperfect correction of unresolved features \citep{Bailey2007}.

An alternative approach is the simultaneous modeling of both telluric and spectral features from the data. 
As the Earth's motion around the barycenter of the Solar System induces a Doppler shift considerably larger than both the motion of telluric features and the size of a single pixel on the detector, these two spectra can be disentangled.
This process is well-established in the analysis of binary star systems through the development of linear models \citep[e.g.][]{Simon1994} and in a Gaussian process framework \citep{Czekala2017}.
In these cases, both spectra are assumed to be unchanging in time, which is a reasonable approximation of a stellar spectrum but not necessarily of the telluric spectrum.
A more complicated model with time variability in the telluric spectrum may provide a more accurate fit. 
Work by \citet{Artigau2014} demonstrates that a Principal Component Analysis (\acronym{PCA}) approach is an effective way of parameterizing telluric spectral variability as a low-dimensional model derived from observational data. 
\citet{Artigau2014} use a library of telluric standard observations, which requires a significant investment of observing time to build up, but in principle such a data-driven model should be possible to derive from typical stellar observations if the same star is observed many times at different barycentric \RV shifts.

Just as imperfect telluric modeling can be a noise source in \EPRV analyses, the choice of stellar template can also be a major source of error. 
For stabilized, non-gas-cell \RV spectrographs, a standard approach has been to adopt a quasi-binary mask consisting of weighted top-hat functions at the expected locations of informative stellar absorption lines and cross-correlate this mask with the observed spectrum \citep[e.g.][]{Baranne1979, Pepe2002}. 
This approach is limited by the accuracy of the mask, and since most masks are built for a broad category of spectral type rather than customized for the individual star in question, it is unlikely that this technique retrieves maximally precise \RVs. 
Deriving a custom spectral template by stacking all spectra iteratively as the \RVs are determined has been shown to be a superior approach for stars with complex spectra \citep{AngladaEscude2012, Zechmeister2018}. 

Data analysis pipelines for absorption cell instruments have long used a stellar template that is customized to the star in question, although traditionally this template is \changed{derived from} a single observation taken for this purpose \citep[e.g.][]{Butler1996}. 
Higher quality templates can be derived iteratively from the data \citep{Sato2002, Gao2016}. 
Such data-driven templates have the benefit of being both customized to the star in question and high-\SNR without requiring observing overhead. 
\changed{Moreover, these templates contain valuable scientific information, as an optimally-combined stellar spectral template will yield the most precise possible constraints on the spectroscopic parameters and abundances for the star.}

Common to the issues of telluric correction and stellar template building is the fact that all of the necessary information about the unknown spectra is encoded in the data. 
With a simple data-driven model, one could learn both stellar and telluric spectra simultaneously with the stellar \RVs. 
Here we develop and implement such a data-driven model to infer the telluric and stellar spectra and calculate the stellar \RV at each observed epoch. 
The telluric model component may vary with time in a low-dimensional manner, which is also inferred from the data. 
Our model requires no prior knowledge of spectral features for the star or for the Earth's atmosphere. 
As such, it does not yield absolute measurements of \RVs, only highly precise relative measurements between epochs. 

Our approach \changed{to the \EPRV problem} is similar to that taken by \citet{Gao2016}, but differs in that we take the stellar spectrum as a latent variable to be optimized rather than iteratively determining it by stacking model residuals. 
\changed{Similar approaches of learning template spectra for stars and tellurics from the data directly have been used in the literature, although they have not previously been demonstrated to achieve precise stellar \RVs \citep[e.g.][]{Hadrava2004, Hadrava2006}.}

In this work, we focus on the ultra-stabilized spectrograph case, i.e. a reliable instrumental calibration and no absorption cell. 
We also assume that multiple epochs of observations exist and that these epochs are spread out across the observing season(s). 
This assumption is necessary to enable the disentangling of telluric features from the stellar spectrum. 
In this sense our method is intended as a post-processing step, not a real-time data reduction service. 
However, the implementation that we present here is designed for flexibility and easy extensibility, and we discuss potential ways to overcome these limitations.

In Section \ref{s:methods}, we outline the model and its key underlying assumptions. 
We present an open-source implementation of this method in \python and \TF called \wobble.
In Section \ref{s:results}, we demonstrate \wobble's capabilities by applying our method to \HARPS archival data for three target stars: the canonical planet-hosting solar analog 51 Peg, the quiet M dwarf \Mdwarf, and the Hot Jupiter host HD 189733. 
We look further into the detailed time-variable telluric spectra inferred from these data in Section \ref{s:tellurics}. 
We revisit many of the assumptions underlying \wobble in Section \ref{s:future} and outline potential ways of adapting \wobble\ for such cases as instruments with absorption cells, intrinsic time variability in the stellar spectrum, and lower-quality data. 
Finally, we conclude with a brief summary in Section \ref{s:conclusion}.

\section{Methods}
\label{s:methods}
\subsection{Model Assumptions}
\label{s:assumptions}

The model underpinning \wobble is designed to be flexible and easily extensible to a variety of situations. 
However, a few assumptions are made in this work to simplify the implementation, and we outline those here. 
Many of these assumptions could be eliminated with relatively straightforward modifications to the method. 
We revisit these in Section \ref{s:future}.

First, we assume that the wavelength calibration and spectral extraction of the instrument are perfect: that is, we begin at the stage of having 1D extracted spectra and corresponding wavelength grids in hand and we do not model any corrections to the wavelength solution. 
\changed{Similarly, we assume that the line spread function of the instrument remains perfectly constant from one exposure to the next. This assumption is needed because it allows us to extract constant instrumentally-broadened template spectra rather than explicitly modeling and solving for time-variable broadening effects. However, as we discuss in Section \ref{s:data-changes}, even within the limits of this assumption extracting already-broadened templates is not strictly correct. We do this for simplicity only and leave the general case to future work.}

We assume that the spectra can be modeled as the product of a finite and fixed number of components.
For the cases shown in this work, two components are used: a stellar spectrum which is invariant in shape but may be Doppler-shifted, and a telluric spectrum which is fixed to the observatory rest frame but varies in shape. 
We choose to work in log(flux) space so that the data to be modeled are simply a sum of the component spectra.

For the case of the telluric component, whose spectrum is allowed to vary with time, we assume this spectral variability is low-dimensional. 
This assumption is physically motivated in the sense that a relatively small number of molecular species contribute to the telluric absorption spectrum. 
It is also needed in a practical sense, since every additional dimension over which the telluric spectrum can vary adds several thousand more free parameters to the model.

We assume that the stellar spectrum is invariant with time. 
This assumption does not hold true in detail and we comment on this in Section \ref{s:model-changes}. 

We assume that both the stellar and telluric spectra are approximately located at zero in logarithmic flux (or unity in linear) with small deviations due to absorption lines. 
As a result of this assumption, we are able to apply L1 and L2 regularization to the spectral templates. 
Regularization is a commonly used technique in machine learning, where large numbers of free parameters are standard. 
It is equivalent to applying a prior to the parameters which pushes them toward zero in the absence of strong evidence otherwise from the data. 
The strength of the regularization may be tuned to suit the data at hand through a cross-validation scheme: for example, the best-suited regularization in the bluest spectral orders may be much stronger for the telluric spectrum, where few features are present, than it is for the star, which generally has a dense forest of spectral lines. 
The exact implementation and validation of regularization in this model is further described in Section \ref{s:model-eqns}.

Aside from this regularization to push model components to zero in the low-\SNR regime, the model makes no assumptions about the shape of the stellar or telluric spectra. 
We solve for spectral templates for each component as a series of control points with no imposed correlations between them, meaning that the line spread function, covariances between lines arising from the same species, and other such physically expected correlations are not built into the model but must be learned in the process of optimizing. 
In addition to keeping this model simple and linear, this means that no physical knowledge about the object being observed is needed to extract its \RVs.

To make our method practically feasible, we assume that the number of observations $N$ in the data set is large ($N \gg N_{components}$) and spread out across the observing season. 
The quantity of spectra needed is a fundamental restriction rooted in the fact that every spectrum being modeled introduces a large number of free parameters to the model. 
In most cases, the epochs of these observations will also need to span a significant fraction of the observing season. 
In order to disentangle the stellar and telluric spectral components, the spectra must undergo Doppler shifts with respect to each other that are at least as large as a resolution element of the spectrograph. 
For stars which do not physically undergo large \RV shifts over very short timescales, this means that it is necessary to observe the star over a significant fraction of the year to take advantage of the changes in the Earth's projected motion. 
We will comment on ways to overcome these restrictions in Section \ref{s:future}.

Finally, we assume that, when the stellar and telluric components are properly optimized, any remaining noise may be approximated as Gaussian. 

\subsection{Model Specification}
\label{s:model-eqns}

We take the data to be the $M \times N$ matrix \mmatrix{y}, where each entry $y_{m,n}$ is the logarithm of the observed flux for pixel $m$ of $M$ at epoch $n$ of $N$. 
We also have a corresponding $M \times N$ matrix of wavelength solutions \changed{which we call \mmatrix{\xi}, where each entry $\xi_{m,n}$ is the logarithm of the wavelength for pixel $m$ at epoch $n$}.

\changed{For each data column \mvec{y_n}, our model prediction \mvec{f_n} can be treated as the sum of stellar and telluric contributions at time $n$:
\begin{equation}
\mvec{f_n} = \mvec{f_{\star,n}} + \mvec{f_{t,n}} + \textrm{noise}.
\end{equation}}

\changed{The stellar spectrum contribution is: 
\begin{equation}
\mvec{f_{\star,n}} = \mmatrix{P}(\mvec{\xi_n}, \mvec{\xi_{\star}}, v_{obs,n}) \cdot \mvec{\mu_{\star}},
\label{eqn:star}
\end{equation}
where \mvec{\mu_{\star}} is a spectral template of log-fluxes and 
\mvec{\xi_{\star}} is the corresponding vector of template log-wavelengths. 
The exact values of \mvec{\xi_{\star}} can be chosen somewhat arbitrarily under the conditions that the grid has uniform spacing $\Delta\xi_{\star}$, covers the entire wavelength range of the data, and is over-sampled with regards to the observed spectrum's wavelength grid.
\mmatrix{P} is a linear operator whose function is to apply a Doppler shift by observed velocity $v_{obs,n}$ and
interpolate \mvec{\mu_{\star}} from the \mvec{\xi_{\star}} template grid to the \mvec{\xi_n} data grid.
Each entry of the \mmatrix{P} matrix can be defined by a sum of weighted indicator functions (where an indicator function is denoted here as $\mathbb{1}(x)$ and is defined to have value 1 when condition $x$ is fulfilled and zero otherwise):
\begin{equation}
\begin{split}
 P_{i,j} =  &\ \bigg(\frac{\xi_{n,i} - \xi'_{\star,j}}{\Delta\xi_{\star}}\bigg) \cdot \mathbb{1}\bigg(0 \leq \frac{\xi_{n,i} - \xi'_{\star,j}}{\Delta\xi_{\star}} < 1\bigg) \\
  &+ \bigg(1 - \frac{\xi_{n,i} - \xi'_{\star,j}}{\Delta\xi_{\star}}\bigg) \cdot \mathbb{1}\bigg({-1} < \frac{\xi_{n,i} - \xi'_{\star,j}}{\Delta\xi_{\star}} \leq 0\bigg), \\
 \end{split}
\end{equation}
where \mvec{\xi'_{\star}} is the Doppler-shifted template grid:
\begin{equation}
 \mvec{\xi'_{\star}}(v) = \mvec{\xi_{\star}} + D(v) \equiv \mvec{\xi_{\star}} + \frac{1}{2} \ln \left(\frac{1 - v/c}{1 + v/c}\right).
\end{equation}}
Because \mmatrix{P} is quite sparse, in practice we do not instantiate the full matrix when performing calculations. 
\changed{At this point, it is also useful to note that \mmatrix{P} could, in principle, encode an instrumental line spread function (\LSF). 
We return to this point in Section \ref{s:future}.}

The apparent stellar \RV, $v_{obs,n}$, is a \changed{combination} of the star's actual velocity \changed{in the barycentric reference frame} $v_{\star,n}$ and the projected motion of the Earth about the Solar System barycenter (Barycentric Earth Radial Velocity or \BERV), the latter of which is known. 
\changed{For the purposes of this work, we use the \BERV furnished by the \HARPS pipeline. 
We also make the approximation that the observed velocity is a simple sum of the actual velocity and the \BERV. 
In detail, this approximation incorrectly assumes that the transverse component of the Earth's motion with respect to the target star is negligible, which may affect our resulting \RV precision by up to a few \ms \citep{Wright2014, Wright2019}. Nevertheless, because only the 1D \BERV is supplied by (and presumably used by) the \HARPS pipeline, we adopt this approximation to ensure that our results are comparable to those reported by the pipeline.}

The telluric spectrum contribution is:
\changed{\begin{equation}
\mvec{f_{t,n}} =  a_n(\mvec{\mu_{t}} + \mmatrix{W_{t}} \cdot \mvec{z_n}).
\end{equation}}
In addition to its mean spectrum \mvec{\mu_t}, the telluric component also includes a time-dependent term assembled from two variables: \mmatrix{W_t}, a matrix of ``basis vectors'' for the span of telluric spectral variations, is weighted by \mvec{z_n} to form the spectral contribution at epoch $n$. 
\changed{\mmatrix{W_t} has the shape $M' \times K$ and \mvec{z_n} is a $K$-vector, where $M'$ is the length of some vector of template wavelengths \mvec{\xi_{t}} and $K$ is the number of basis vectors used.\footnote{While we refer to these as basis vectors for simplicity, note that they are not actually constrained to be orthogonal.}} 
For the purposes of this work we found good performance with $K$ set to 3, but this may vary for other applications. 
\changed{The template wavelength grid \mvec{\xi_{t}} should have similar properties to \mvec{\xi_{\star}}.} 
Finally, the net telluric spectrum (mean + time-variable components) is weighted by the airmass at the time of observation $a_n$, a known quantity.

\changed{The contribution of each data epoch to the net log-likelihood may be evaluated as
\begin{equation}
\ln \mathcal{L}_n = -\frac{1}{2} (\mvec{y_{n}} - \mvec{f_{n}})^T \mvec{C_{n}}^{-1} (\mvec{y_{n}} - \mvec{f_{n}}),
\label{eqn:lnlike}
\end{equation}
}with \mmatrix{C_n} representing the covariance matrix of uncertainties on the data. 

The number of free parameters in this model is large: we must optimize every grid point in the mean spectral templates and telluric basis vectors along with the stellar \RV and the telluric basis weights for each epoch. 
We deal with potential over-fitting issues by applying L1 and L2 regularization to the spectral templates, as discussed in Section \ref{s:assumptions}. 

L1 normalization adds a term to the log-likelihood that takes the form:
\begin{equation}
\Lone{\lambda}{p} \equiv -\lambda \sum_{i} | p_{i} | ,
\end{equation}
where \mvec{p} is the vector of parameters to be normalized (in this case \mvec{\mu_{\star}}, \mvec{\mu_{t}}, or \mmatrix{W_{t}}) and $\lambda$ is the regularization amplitude.
Similarly, L2 normalization adds a term of the form:
\begin{equation}
\Ltwo{\lambda}{p} \equiv -\lambda \sum_{i} p_{i}^2 .
\end{equation}

The effectiveness of the regularization depends sensitively on the value of $\lambda$ used: if $\lambda$ is too high, real features will be lost as the parameters are forced to zero, whereas setting $\lambda$ too low will make the regularization ineffective, leaving the model vulnerable to overfitting. 
We set regularization amplitudes for \wobble using a cross-validation scheme. 
In brief, we randomly select $10-15$\% of the total epochs to set aside as a validation set and, using some value of $\lambda$, run the model optimization on the remaining epochs (the training set). 
The resulting best-fit spectral templates and basis vectors are taken as fixed and the time-dependent terms only (\RVs and basis weights) are optimized for the validation epochs. 
The $\chi^2$ for the validation epochs can then be adopted as a goodness-of-fit measurement for the $\lambda$ value, and the procedure is repeated for different $\lambda$s to choose the best regularization amplitude. 
In theory, we might wish to regularize using an optimization metric based on \RV accuracy rather than $\chi^2$ over pixels; however, the lack of known ``ground truth'' in the \RV behavior of stars makes this not currently feasible. 

Since we have multiple regularization amplitudes to set, we begin by hand-setting all amplitudes to a reasonable starting guess and optimize each amplitude sequentially with cross-validation. 
Generally speaking, the L2 regularization tends to be stricter than L1 and the telluric components are more strongly regularized than the stellar components, so we aim to go roughly from most to least sensitive regularization component when tuning the amplitudes.
We found that good performance came from tuning the L2 regularization amplitudes for the mean telluric and stellar spectra, followed by the L1 regularization amplitudes for the same, followed by L2 and L1 for the time-variable telluric basis vectors. 

With regularization included, the final model likelihood to be optimized is:
\begin{equation}
\begin{split}
\ln \mathcal{L} = & -\frac{1}{2} \sum_{n} (\mvec{y_{n}} - \mvec{f_{n}})^T \mvec{C_{n}}^{-1} (\mvec{y_{n}} - \mvec{f_{n}})  \\
 & + \Lone{\lambda_1}{\mu_{\star}} + \Ltwo{\lambda_2}{\mu_{\star}}   + \Lone{\lambda_3}{\mu_{t}} + \Ltwo{\lambda_4}{\mu_{t}} \\
 &  + \Lone{\lambda_5}{W_t}  + \Ltwo{\lambda_6}{W_t} + \Lone{1.0}{z} ,
\end{split}
\end{equation}
where the regularization amplitude on the basis weights is arbitrarily set to unity, and all other regularization amplitudes $\lambda$ are set by grid searches using the above-described validation procedure.
The basis-weight regularization amplitude can be set arbitrarily to unity because there is a perfect degeneracy between amplifying the basis vectors and attenuating the basis weights.
This regularization choice breaks that degeneracy.
The basis weights are regularized to encourage our desired outcome where the mean telluric spectrum contains as much telluric information as possible, while the variable basis picks up only the necessary time-variable changes.

\subsection{Optimizing the Model}
\label{s:optimizing}

As an initial guess, we set the star to be stationary, e.g. $v_{obs} = $ \BERV at all epochs. 
We may then initialize the stellar spectrum \mvec{\mu_{\star}} by Doppler-shifting the data and calculating the median flux across \BERV-corrected spectra in bins at each model wavelength \changed{\mvec{\xi_{\star}}}. 
The telluric template is initialized similarly by using the residuals after the stellar contribution has been removed, again binning in model wavelength (this time without applying any Doppler shift to the spectra) and taking the median values of each bin. 
Finally, we initialize the telluric spectrum's basis vectors \changed{\mmatrix{W_t}} and weights \changed{\mvec{z_t}} by performing Principal Component Analysis (\acronym{PCA}) on the residuals after both the stellar spectrum and the mean telluric spectrum have been removed. 
The $K$ highest eigenweights and their corresponding eigenvectors are taken as \changed{the starting guess for} the basis weights and vectors. 

Once all parameters are initialized, we optimize them iteratively. 
The likelihood function is maximized first by varying the stellar and telluric templates (including the telluric basis vectors). 
This step is a convex optimization, meaning that a global optimum for the template parameters should be reached under the condition of fixed time-dependent variables. 
Next, the templates are held fixed and the time-dependent parameters (stellar \RVs and telluric basis weights) are varied to maximize the likelihood function again. 
Technically the velocities are location parameters and their optimization is not convex.
However, once the velocities are known to a small fraction of a pixel (which they usually are in practice, given an accurate estimation of the \BERV), the linearized problem becomes convex at each iteration.
We repeat this procedure, optimizing spectra and time-dependent parameters in turn, until the likelihood appears to converge, typically within 100 iterations.
This iterative procedure is equivalent to a full simultaneous optimization in the limit of many iterations to convergence.
After convergence, we estimate the uncertainties on the parameters by
approximating the likelihood function near its maximum as Gaussian where the
covariance matrix is the negative inverse Hessian (or second derivative
matrix) of the log likelihood function with respect to the parameters.

\subsection{Combining Spectral Orders}
\label{s:combining-orders}

Most \EPRV instruments are echelle spectrographs spanning many orders. 
These orders are often treated as independent spectra when extracting \RVs, and the \RVs for each order are then combined in some manner to get a final time series. 
We follow this precedent and optimize the \wobble model individually for each order. 

After obtaining an $N$-epoch set of observed stellar \RVs from each of the $R$ spectral orders, we combine them by modeling each \RV $v_{n,r}$ as a combination of time-dependent stellar \RV, order-dependent \RV offset, and a Gaussian noise term:
\begin{equation}
v_{n,r} = v_n + v_r + \mathcal{N}(0,\,\sigma_{n,r}^{2} + \delta_r^2 )\,,
\end{equation}
where $v_n$ and $v_r$ are the characteristic \RVs at epoch $n$ and order $r$, $\sigma_{n,r}$ is the estimated measurement uncertainty on $v_{n,r}$, and $\delta_r$ is an additional jitter term specific to order $r$. 
The order-dependent \RV offset should be small and indeed generally is consistent with zero in the solutions.

\subsection{Implementation}
\label{s:implementation}

The above-described model can be implemented in a variety of ways. We chose to build our code, \wobble, in \code{python} using \TF \citep{Abadi15}. 
\TF is a model building framework that has been primarily designed for machine
learning applications, but at its core \TF is fundamentally a collection of
highly optimized routines for doing linear algebra and efficiently computing the
derivatives of these models with respect to large numbers of parameters.
\wobble can be represented in this framework so we have re-purposed \TF for our
needs.
By using this framework, we benefit from the high performance and scalability
of the implementation, as well as the algorithms implemented within \TF for
fitting large numbers of parameters to large datasets.
The necessary optimizations over many parameters can be performed with high efficiency by \TF: the below-described analysis of 91 \HARPS spectra of 51 Peg, including the optimization of 72 spectral orders, runs in 60 minutes on a standard Mac desktop.

Our code is made open-source under the \acronym{MIT} license and is publicly available on GitHub.\footnote{\url{https://www.github.com/megbedell/wobble}; in this work, we use the version released as Zenodo v0.1.0 \citep{wobble-code}}

\section{Application to \HARPS Data}
\label{s:results}

For the purposes of this work, we chose to test the performance of \changed{the \wobble method} on archival spectra from the High Accuracy Radial Velocity Planet Searcher (\HARPS) spectrograph \citep{Mayor2003}. 
\HARPS has operated continuously since 2003 and as such has an extensive catalog of publicly available data. 
Furthermore, its excellent instrumental stability and precise calibration make the data ideally suited to our method, which relies on having an accurate wavelength solution for every spectrum. 

All data were obtained from the \acronym{ESO} public data archive in the form of ``\code{e2ds}'' spectra.\footnote{Based on observations made with ESO Telescopes at the La Silla Paranal Observatory under programme IDs 091.C-0271, 183.C-0437, 191.C-0505, 072.C-0488, 089.C-0497, 099.C-0880, and 60.A-9700.} 
These data come as extracted \changed{order-by-order} 1D spectra in blocks of 72 orders by 4096 pixels per order.
The airmass of the observation and the calculated \BERV are both provided in the \acronym{FITS} header of the spectral data file. 
We obtained wavelength solutions for each spectrum from the dedicated \HARPS calibration archive, also maintained by \acronym{ESO}.

Before running the above-described model optimization on the data, we first mask out any unreliably measured spectral regions and do a continuum normalization on each spectrum. 
This masking is done by setting the inverse variance on the masked data point to zero, effectively removing it from the fit. 
In brief, we mask pixels whose extracted flux is below zero as well as regions at the edges of the spectral orders where the local \SNR falls below 5. 
After masking these data, we convert the flux to log space and continuum normalize by fitting and subtracting a polynomial to an asymmetrically clipped subset of the data. 
\changed{We found good results from clipping all pixels outside of the range [$-$0.3$\sigma$, $+$3$\sigma$] as a way of effectively removing the absorption features; fitting a sixth-order polynomial; and iteratively repeating with the pixel clipping set by the residuals to the previous fit until the selection of clipped pixels is stable. 
This was done independently for each echelle order.}

To each pixel in the normalized, logarithmic spectrum, we assign an uncertainty variance (squared error).
Technically this uncertainty is assumed to be Gaussian in the log space, which is an incorrect approximation, but not very far off for high signal-to-noise data. 
For each order we assign to each log flux an uncertainty variance that is the inverse of the mean signal-to-noise-squared of the pixels in that order (as reported by the \HARPS pipelines), scaled up or down by the raw linear flux observed in that pixel in the data prior to logging and normalizing. 
This is the best Poisson estimate we can make given that we are working in the log space, and we don't have individual-pixel uncertainty estimates. 
Technically a small bias is introduced by taking the logarithm of the flux, but this bias is very small at the relevant signal-to-noise ratios; furthermore, this small bias \changed{is in the flux direction and will not necessarily map onto the radial velocities obtained.}

If possible, the final stellar \RV retrieved by \wobble is corrected for intra-night drift in the wavelength solution by subtracting off an \RV drift term specified in the data header, consistent with the standard \HARPS pipeline. 
This drift is calculated from the simultaneous reference lamp and is therefore only available in certain observing modes. 
Its contribution to the \RV solution is usually below 1 \ms, supporting our assumption that the wavelength solution provided by the \HARPS pipeline is generally accurate.

Below we describe the results of applying this procedure and optimizing the \wobble model using \HARPS data for stars of different types: a G dwarf with a known planet, an \RV-quiet M dwarf, and an early-K dwarf undergoing a planetary transit event.

\begin{figure*}[ht!]
\centering
\includegraphics[width=5in]{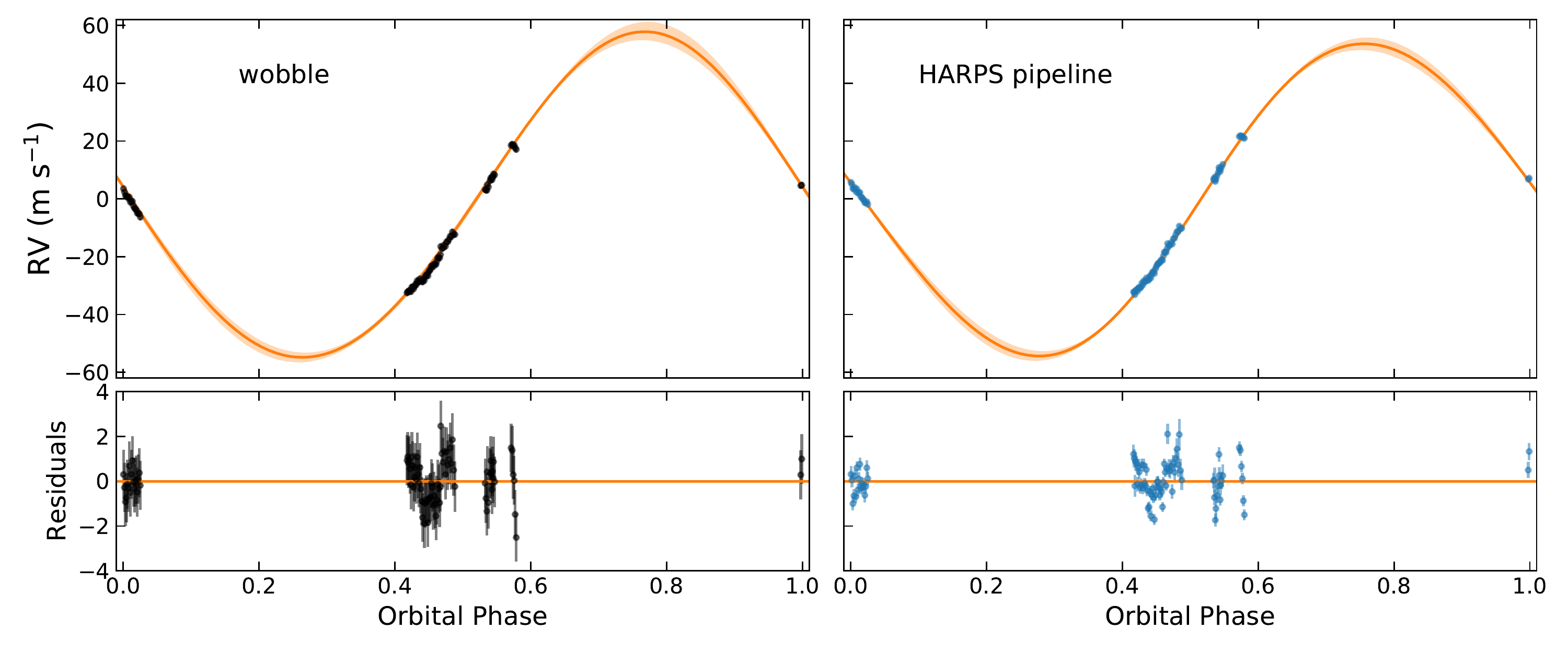}
\caption{Phased orbit of the Hot Jupiter 51 Peg b as recovered from the \wobble\ \RVs (left) and from the \HARPS standard pipeline (right). While the phase coverage of \HARPS observations is sparse, we nonetheless recover orbital parameters consistent with those found by other \RV campaigns. Maximum-likelihood Keplerians (solid orange lines) and 1-$\sigma$ posterior constraints (shaded orange regions) for each data set are overplotted.}
\label{fig:51peg_planet}
\end{figure*}

\begin{deluxetable*}{cccc}
\tablehead{
\colhead{Parameter} & \colhead{Units} & \colhead{\wobble Fit} & \colhead{\HARPS Pipeline Fit}}
\tablecaption{Orbital Parameters of 51 Pegasi b}
\label{tbl:51peg}
\startdata
$K$ & m s$^{-1}$ & $55.57^{+2.28}_{-2.04}$ & $53.84^{+1.96}_{-2.20}$ \\
$P$ & days & $4.2292^{+0.0003}_{-0.0003}$ & $4.2294^{+0.0003}_{-0.0003}$ \\
$t_0$ & JD & $2456546.89^{+0.02}_{-0.02}$ & $2456546.94^{+0.01}_{-0.02}$ \\
$e$ & $-$  & $0.03^{+0.02}_{-0.02}$ & $0.04^{+0.02}_{-0.02}$ \\
$\omega$ & rad & $0.45^{+0.62}_{-1.00}$ & $-1.79^{+0.34}_{-0.32}$ \\
$s$ & m s$^{-1}$ & $0.01^{+0.08}_{-0.01}$ & $0.74^{+0.06}_{-0.07}$ \\
\enddata
\end{deluxetable*}

\subsection{51 Pegasi}

For \wobble's first test, we chose the first known exoplanet host: 51 Pegasi. 
This target is a Sun-like star hosting a planet with an orbital period of 4 days and a mass of 0.5 Jupiter masses \citep{Mayor1995}. 
Its canonical status as the first exoplanet discovered means that large amounts of data exist for this system. 
In particular, archival \HARPS data exist mainly from efforts to observe reflected-light spectra of the planet \citep{Martins2015}. 

We ran \wobble on these archival data to test its performance on recovering a planetary signal with well-known orbital characteristics.
The 91 publicly available spectra in the \HARPS archive are largely concentrated on a few nights of intensive observing, but these nights are sufficiently spread out throughout the year for a wide enough range in \BERV to disentangle the stellar and telluric spectra. 
Their \SNR is generally high, ranging from 100 up to 300 pix$^{-1}$ at the central wavelength regions of \HARPS.

\begin{figure*}[t]
\centering
\includegraphics[width=5.3in]{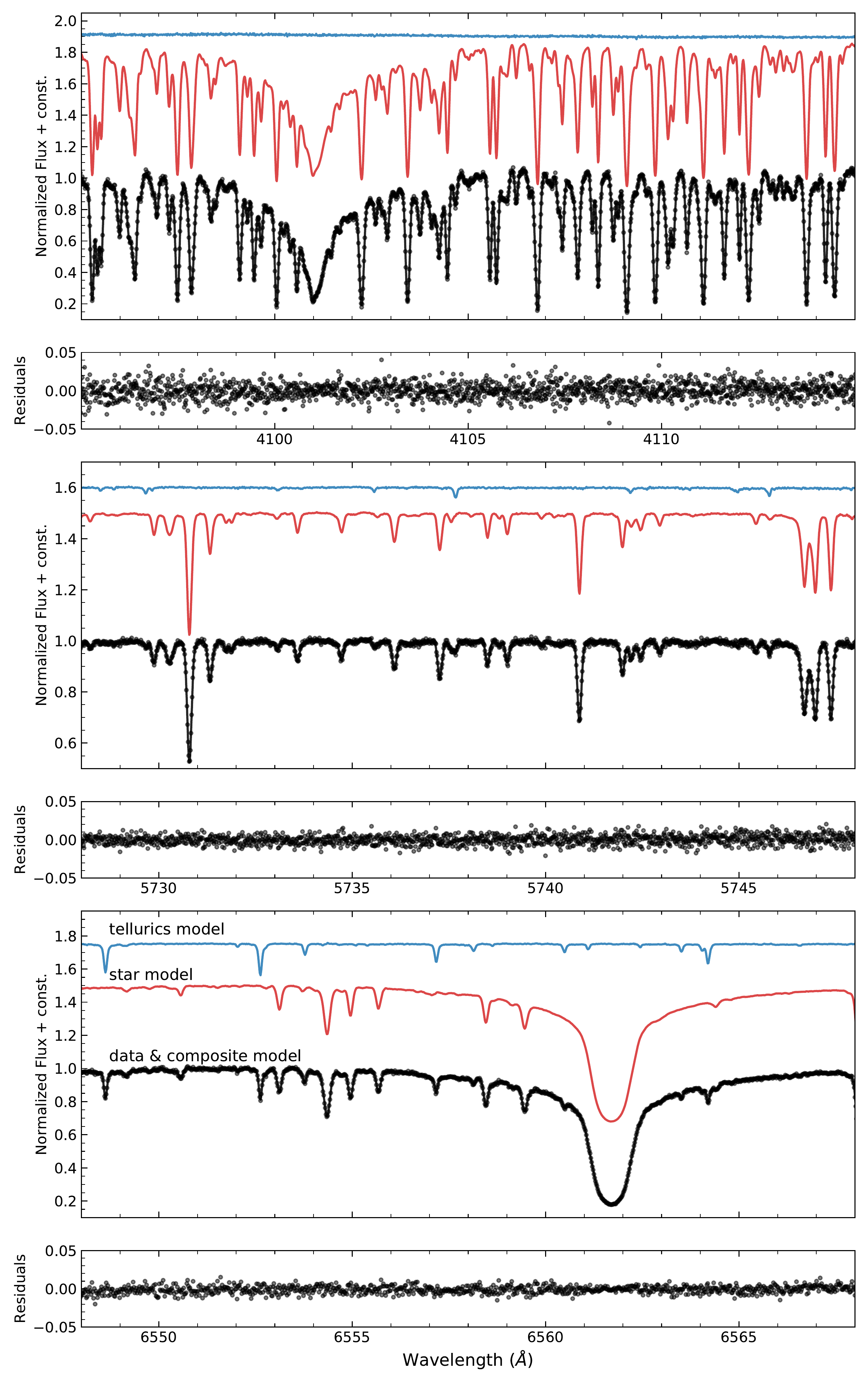}
\caption{Three example echelle orders at a randomly chosen epoch from the \HARPS observations of 51 Pegasi. \changed{Data and best-fit models are plotted in black, while the predicted stellar and telluric spectral contributions from the model fit are plotted in red and blue with arbitrary flux offsets for clarity.} Even in the presence of a strong stellar absorption feature such as the H$\alpha$ line (seen at 6561.5 \ang due to the star's Doppler shift at the plotted epoch), small telluric features are clearly recovered. Residuals after subtracting both star and telluric models are shown below each spectrum.}

\label{fig:51peg_spectrum}
\end{figure*}

Despite the sparse phase coverage available across the planetary orbit, we recover a signal at the expected period and semi-amplitude in the \RVs (Figure \ref{fig:51peg_planet}). 
We fit a Keplerian signal with seven free parameters (period $P$, \RV semi-amplitude $K$, eccentricity $e$, argument of periastron $\omega$, time of periastron $T_0$, \RV offset $c$, and \RV jitter $s$) using the \code{exoplanet} package \citep{exoplanet}. 
The resulting best-fit parameters for the two data sets are generally consistent (Table \ref{tbl:51peg}), and both are comparable to literature values \citep{Mayor1995, Naef2004, Butler2006}. 
The \RV uncertainties derived by \wobble appear accurate based on the negligible jitter in the fit, while the photon-noise-based \RV errors provided by the \HARPS pipeline are much smaller and require a significant jitter to achieve a good fit. 
These results confirm that \wobble is able to extract \RVs with similar precision to the closed-source \HARPS pipeline for Sun-like stars.

Fits to an individual spectrum are shown in Figure \ref{fig:51peg_spectrum}. 
We emphasize that no \textit{a priori} information on e.g. expected spectral line positions and depths were used. 
The \wobble algorithm as it is currently implemented treats each control point of the template spectra as independent and has no line shape parameterization included. 
Nevertheless, the optimized stellar templates clearly reproduce the expected appearance of the spectra in a variety of regimes, from a crowded-line region to a sparser, continuum-dominated region to an extremely strong absorption line like the H$\alpha$ feature. 
Moreover, telluric absorption features are recovered down to a very small amplitude.

It is worthy of note that the stellar spectral model shown in Figure \ref{fig:51peg_spectrum} is an extremely high-\SNR spectrum of 51 Peg. 
\changed{The uncertainties on this spectrum can be estimated from the negative inverse Hessian, similarly to the \RVs. Doing this calculation for a subset of the template in the continuum around 6000 \ang indicates an approximate \SNR of 3000 per 0.02-\ang template ``pixel,'' which is consistent with the expected net \SNR from co-adding each individual \HARPS spectrum. }
\changed{In essence, this template is a time-averaged and telluric-cleaned composite stellar spectrum.} 
Both it and its residuals as a function of time are scientifically valuable outputs of the \wobble method. 
\changed{The template, with its exquisite \SNR, can be used for very precise stellar characterization and abundance analysis. Meanwhile, the time-series residuals contain information about the stellar spectrum's non-Doppler-shift variations in time, a matter of key importance to both stellar physics and the effort to mitigate stellar ``noise'' in \EPRV measurements.}

\subsection{\Mdwarf}
\label{s:Mdwarf}

\begin{figure*}
\centering
\includegraphics[width=5in]{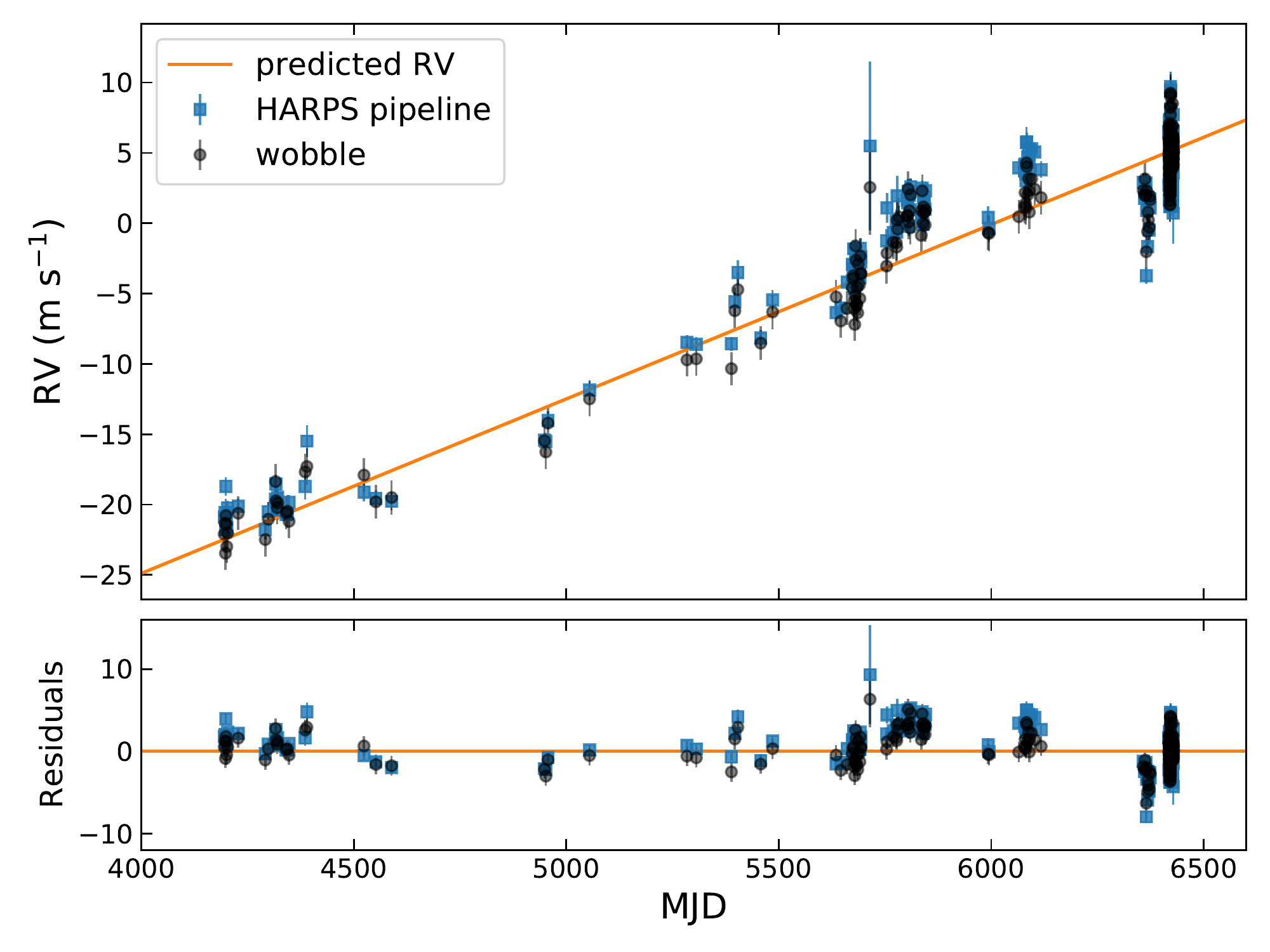}
\caption{Radial velocity measurements for \Mdwarf from \wobble (black circles) and from the standard \HARPS pipeline (blue squares). The median \RV has been subtracted from each data set. The predicted secular change in \RV due to \Mdwarf's projected motion, calculated using \gaia properties, is shown as a solid orange line (top panel). Residuals away from the predicted trend are plotted in the lower panel.}
\label{fig:barnards_rvs}
\end{figure*}

Next, we tested \wobble with 237 epochs of \Mdwarf spectra. 
We did not utilize the additional spectra taken after the \HARPS 2015 optical fiber upgrade, as these would need to be treated as an independent data set \citep{LoCurto2015}. 
\Mdwarf is a mid-M dwarf, and as such its output is low in the optical; the typical \SNR of the spectra in the central wavelength regions of \HARPS ranges from 20-50. 
In practice, this means that the 7 bluest echelle orders were consistently masked and dropped from the fit in accordance with the \SNR criteria outlined previously. 
Regardless, a large amount of data remained, allowing us to test \wobble's performance on observations with substantially different properties from the previous case.

\Mdwarf is one of the nearest stars to us at a distance of less than 2 parsecs. 
It is also the star with the highest known proper motion. 
Its trajectory translates to a projected secular change in \RV of approximately 4.5 $\mathrm{m}~\mathrm{s}^{-1}~\mathrm{yr}^{-1}$ \citep{Kurster2003}. 
This trend has been observed in some data sets, including long-term \HARPS and \acronym{HIRES} observations, although \acronym{UVES} \RVs were inconsistent with the predicted linear slope \citep{Kurster2003, Bonfils2013, Choi2013, Montet2014}. 
Aside from this linear trend, \Mdwarf is commonly used as an \RV standard M dwarf because little stellar activity has been observed and no planets discovered until very recently despite considerable \RV monitoring \citep{Ribas2018}. 

The \RVs found by \wobble are in excellent agreement with the predicted secular motion (Figure \ref{fig:barnards_rvs}). 
Using \gaia parallax and proper motion measurements and following the calculations outlined in \citet{Kurster2003}, we find a secular trend with a slope of 4.53 $\mathrm{m}~\mathrm{s}^{-1}~\mathrm{yr}^{-1}$, deviating from this linearity by less than 1 $\mathrm{cm}~\mathrm{s}^{-1}~\mathrm{yr}^{-2}$ during the decade of \HARPS observations \citep{gaia2016, gaia2018}. 

After subtracting the secular \RV trend, the residuals have low dispersion, as expected for a quiet star with no planetary signals above $K \sim 1.2~\ms$ \citep{Choi2013, Ribas2018}. 
The \acronym{RMS} scatter among \wobble\ \RVs is 2.0 \ms. 
This compares favorably to the scatter of 2.5 \ms among \RVs produced by the standard closed-source \HARPS pipeline. 

We note that many of the residuals have similar non-zero values in both independently produced RV estimates. 
It is possible that these deviations are a real physical effect in the stellar spectrum, but we caution that approximately \ms-level errors are likely being introduced by the barycentric correction. 
Currently we assume that the \BERV provided by the native \HARPS pipeline is correct, but this will not hold true if the sky coordinates entered by the observer deviate from the actual on-sky location of the target. 
This is especially likely to be the case for high proper motion targets like \Mdwarf. 
Indeed, we do note a strong peak at year-long periods in a Lomb-Scargle periodogram of the residual \RVs. 
Thus the \RV scatter derived in this work is only an approximate upper limit on the true precision achievable by \wobble. 

\begin{figure*}[ht!]
\centering
\includegraphics[width=5.5in]{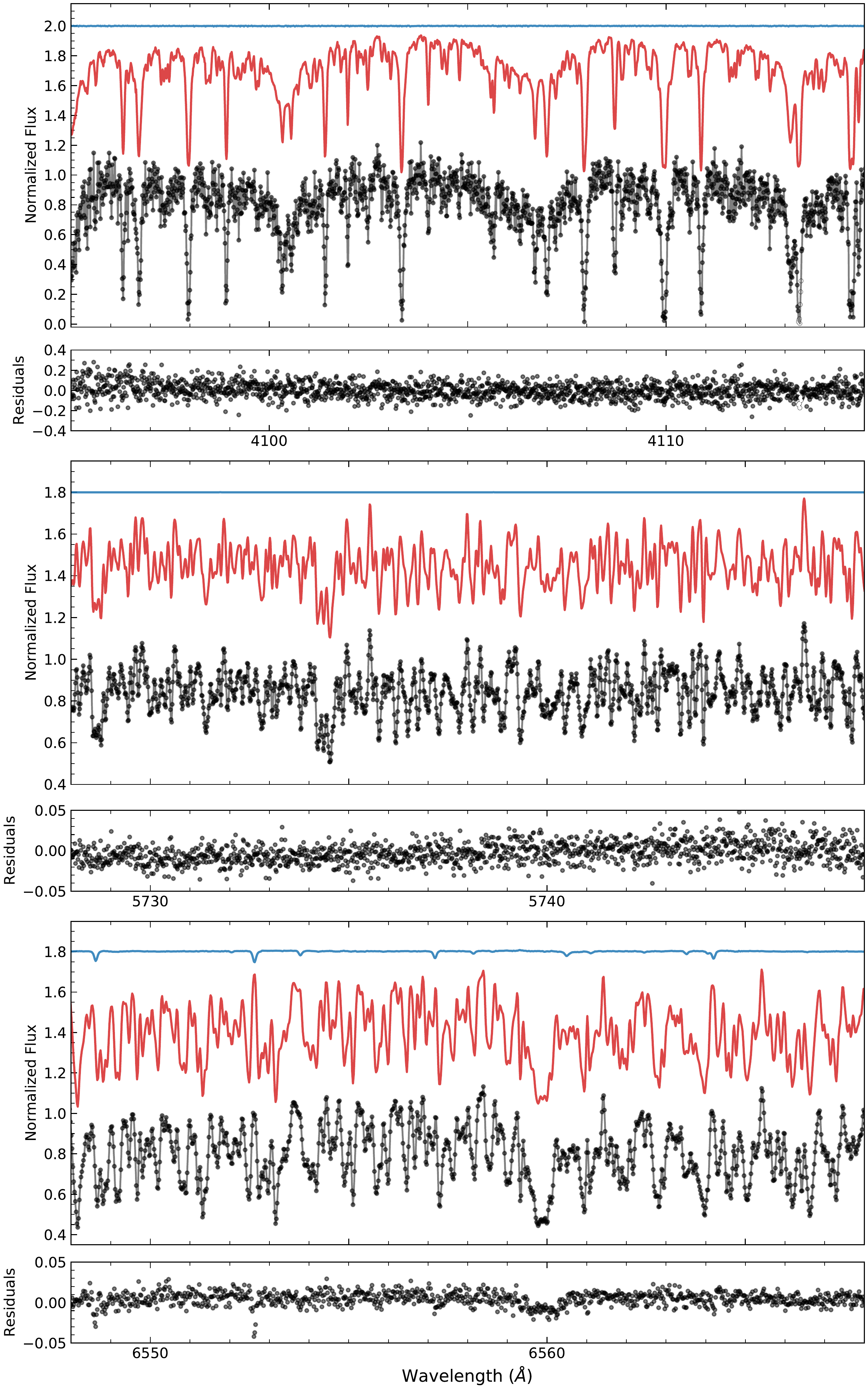}
\caption{Example echelle orders at a randomly chosen epoch from the \HARPS observations of Barnard's Star, plotted as in Figure \ref{fig:51peg_spectrum}.}
\label{fig:barnards_spectrum}
\end{figure*}

\begin{figure*}
\centering
\includegraphics[width=6in]{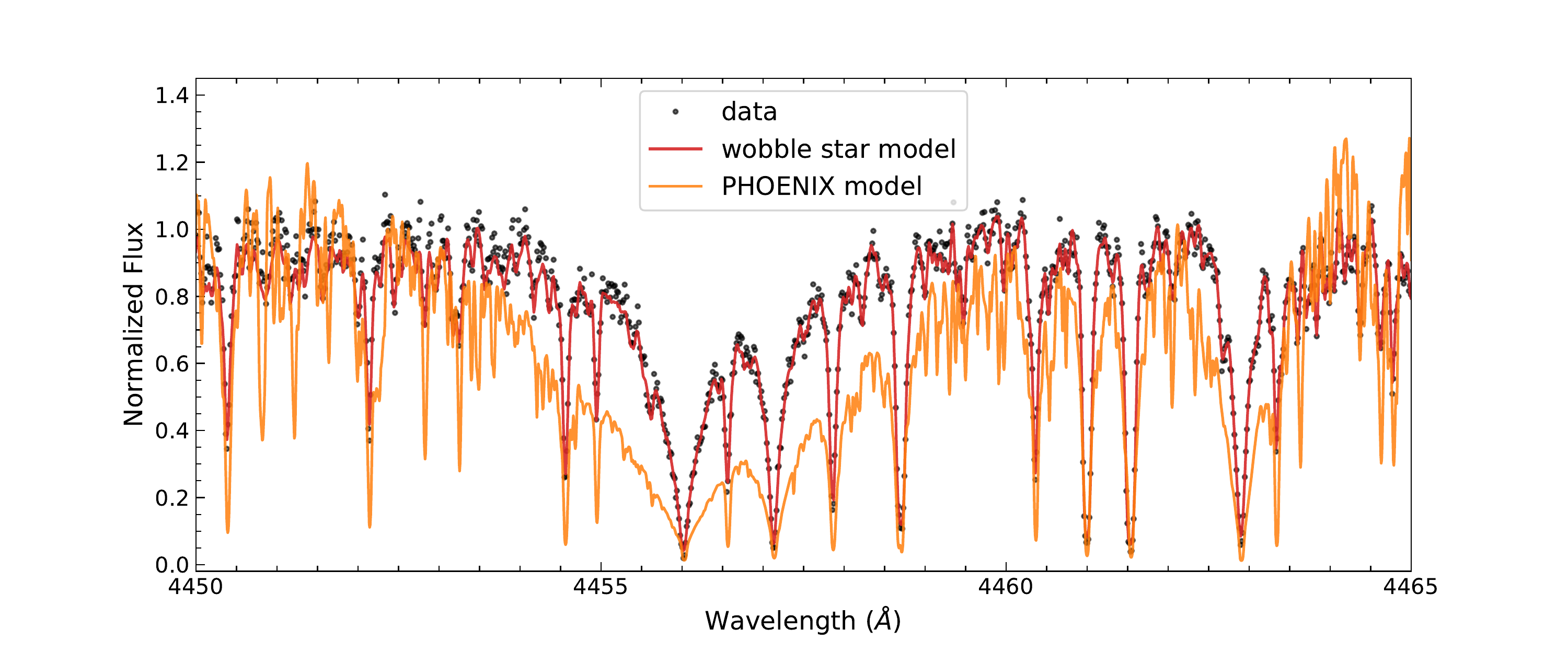}
\caption{A subset of the \Mdwarf spectrum including the sodium doublet at 4456-7 \ang. Data for the highest-\SNR single observation are shown as black points, while the spectral template derived by \wobble from the full data set is shown as a red line. The theoretical \PHOENIX model for an M dwarf with \teff = 3200 K, \logg = 5.0, and \mh = $-0.5$ is shown as an orange line. The \wobble model and data have been Doppler-shifted by eye to match the rest-frame line positions, since \wobble does not deliver an absolute \RV. The data-driven \wobble model of the stellar spectrum shares general strong line locations and relative strengths with the \PHOENIX model, but in detail it delivers a much more accurate fit to the data than the theoretical M dwarf model achieves.}
\label{fig:barnards_model}
\end{figure*}

Testing \wobble on the \Mdwarf data allows us to evaluate its performance in a significantly different regime: the mid-M dwarf spectra are far denser in spectral features than a Sun-like star, and individual observations are at a much lower \SNR than the 51 Peg observations used above, although the total number of spectra is greater. 
The resulting spectral fit is shown in Figure \ref{fig:barnards_spectrum}.

One consequence of working in the low-\SNR regime is that the power to resolve very small telluric features is reduced. 
To keep the template spectra from overfitting the noise, the regularization amplitudes on the templates must be raised by several orders of magnitude relative to what was optimal for the 51 Peg fit above, particularly in orders where no strong spectral features are available. 
This has the side effect of flattening out weak features if no strong lines are present in a given order. 
As a result, we are able to retrieve fewer telluric lines in the bluer orders (compare the middle panels of Figures \ref{fig:51peg_spectrum} and \ref{fig:barnards_spectrum}). 
In redder orders, where the \SNR is higher and more strong telluric lines are present, the features are retrieved (bottom panel of Figure \ref{fig:barnards_spectrum}). 
However, telluric variability was not able to be resolved in the overwhelming majority of orders; in fact, the results presented here were obtained with the tellurics model set to $K=0$ (no variability included). 
We therefore caution that difficult-to-resolve microtellurics and time-variable features may not be reliably disentangled from the stellar spectrum when \wobble is applied to lower signal observations (\SNR $\lessapprox 50$ pixel$^{-1}$).

Despite the extreme noise at the bluer end of the wavelength range, where \Mdwarf is very faint, \wobble successfully retrieves a template spectrum that appears consistent with general expectations for a mid-M dwarf through most of the \HARPS wavelength range. 
As a sanity check, we compare the \wobble stellar template model with a high-resolution \PHOENIX model at \Mdwarf's previously measured spectral parameters \citep[\teff = 3200 K, \logg = 5.0, and \mh = $-0.5$;][and references therein]{Husser2013, Artigau2018}. 
Even in the bluest regions, the placements and relative strengths of the large-amplitude features inferred by \wobble generally compare well with the \PHOENIX predictions (Figure \ref{fig:barnards_model}). 
The \wobble model for \Mdwarf diverges from the theoretical models in the smaller absorption lines and in the degree of broadening for strong lines.  

These results emphasize the potential value of \wobble data products for spectral characterization and model testing, particularly in the case of faint stars and M dwarfs, for which many spectra must be combined to get a reasonably detailed composite spectrum. 
The \wobble algorithm is a simple yet robust method for doing such multi-spectra stacking while preserving telluric-contaminated regions and accounting for unknown \RV shifts.

\subsection{HD 189733}

\begin{figure*}
\centering
\includegraphics[width=6in]{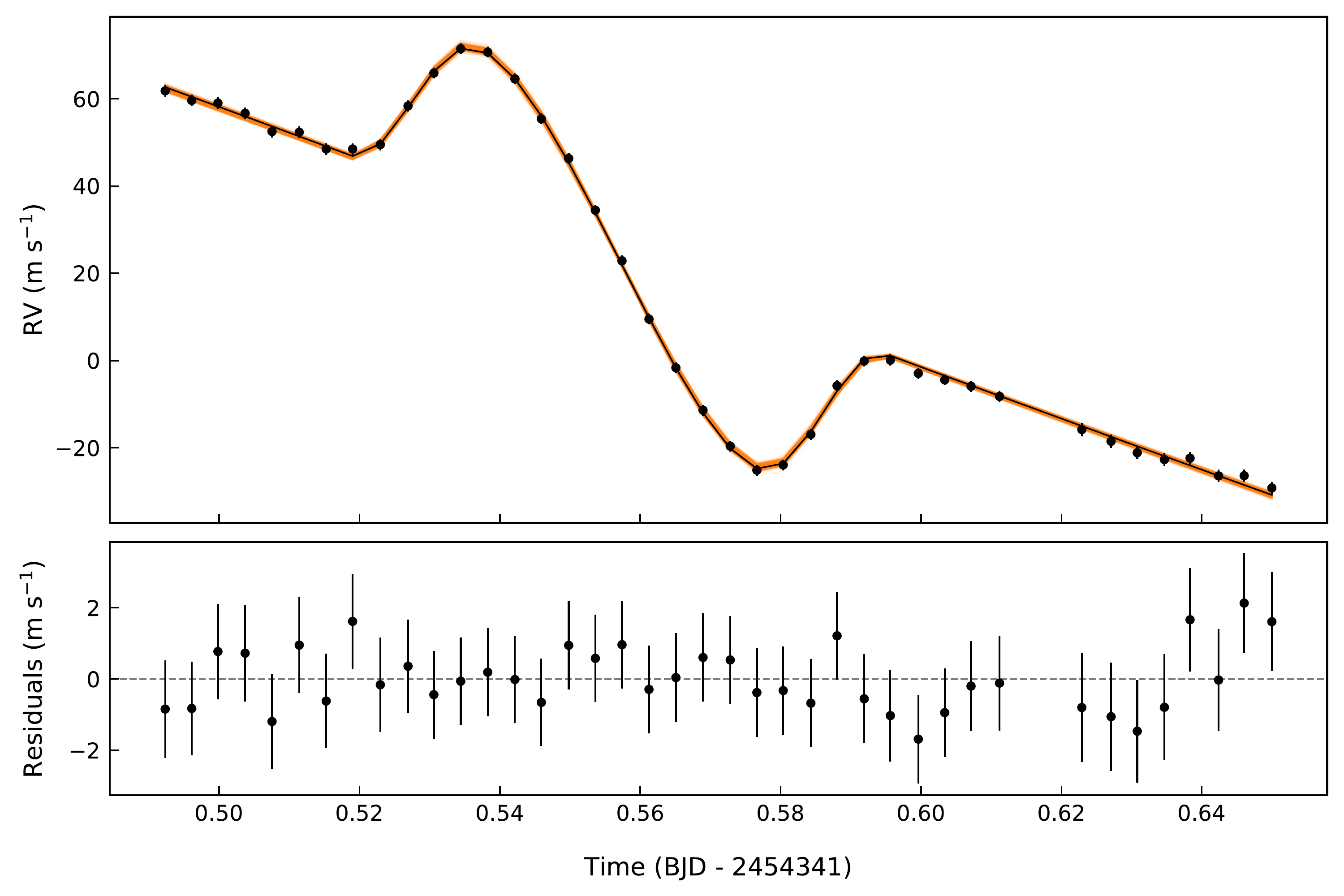}
\caption{\RVs measured by \wobble for a single night of observations of the Hot 
         Jupiter host HD 189733. The \RM is clearly seen as the planet transits 
         the star. The top panel shows the data (black points), the maximum
         likelihood fit using \starry (black line), and 500 posterior samples
         (orange lines). The bottom panel shows the residuals of the maximum
         likelihood model fit.}
\label{fig:hd189_rm}
\end{figure*}

While the majority of \EPRV measurements are made for the purposes of observing stellar reflex motion due to planetary orbits, several other applications to exoplanet characterization exist for such high-precision time-series spectra. 
One of these uses is measuring the \RM, in which the apparent stellar \RV is observed during planetary transit as a way of mapping the stellar surface and learning the spin-orbit inclination of the system \citep[e.g.][]{Queloz2000, Winn2005}. 
While the \RM technically manifests in the stellar spectrum as a distortion to the line profile rather than a true Doppler shift, it is typically detected using similar methods to a standard \RV analysis. 

We apply \wobble to a data set consisting of a single transit of the hot Jupiter HD 189733b to test its sensitivity to stellar line asymmetries as well as its performance in the regime of a single night's observations. 
The observations in question consist of a consecutive series of 40 spectra, each with an \SNR $\sim 90$, taken on the night of August 28, 2007.

Because all observations came from a single night, there will be no significant shift of the stellar spectrum with respect to the telluric spectrum and the power of \wobble to disentangle the two is severely limited. 
For this reason, we fixed the telluric spectrum to a constant template. 
We derived this template by running \wobble on the 51 Peg data with non-time-variable tellurics and adopting the resulting high-quality time-invariant telluric spectrum. 
The stellar spectrum of HD 189733 and its \RVs were left as free parameters.

The resulting \RV signal retrieved by \wobble is shown in Figure \ref{fig:hd189_rm}. 
We fit the signal using \starry \citep{Luger2018}, which computes analytic occultation
light curves for bodies whose surfaces can be decomposed into sums
of spherical harmonics. Since the radial component of the velocity field of a 
differentially rotating star can be expressed in terms of polynomials in 
$x = \sin\theta\cos\phi$, $y = \sin\theta\sin\phi$, and $z=\cos\theta$, where
$\theta$ is the polar angle and $\phi$ is the azimuthal angle \citep[c.f. Equation 91 in][]{Short2018},
this velocity field may be expressed exactly in terms of spherical harmonics
\citep{LugerBedell2019}.
We therefore use \starry to fit for the equatorial rotational velocity, the inclination,
the obliquity, and the shear due to differential rotation of the star, closely
following the analysis performed in \citet{Cegla2016}.
We infer a projected stellar obliquity $\lambda = -0.43 \pm 0.34^\circ$, in close agreement
with the value reported in \citet{Cegla2016}. Our inferred values for the other parameters
are broadly consistent with the results in \citet{Cegla2016}, except with significantly
higher uncertainty due to the fact that we are unable to constrain the stellar inclination
due to the $v\sin i$ degeneracy. We attribute the narrower posteriors in \citet{Cegla2016}
to a difference in the choice of prior.
\footnote{See \href{https://github.com/megbedell/wobble/tree/master/paper/figures/HD189733/DifferentialRotationWithSphericalHarmonics.ipynb}{this interactive notebook} 
          for a derivation of the spherical harmonic decomposition of the stellar velocity field. 
          The notebook used to perform our analysis and produce Figure~\ref{fig:hd189_rm}, along with the full posterior constraints we obtain, can be found 
          \href{https://github.com/megbedell/wobble/tree/master/paper/figures/HD189733/HD189733bWithStarry.ipynb}{here}.}

These results confirm that the \wobble method measures line asymmetries as \RV shifts in the same manner as traditional \RV analysis techniques. 
While this is a useful approximation in the case of the \RM, in general the confusion of these two spectral changes is a major cause of correlated noise in \RV time series \citep[e.g.][]{Queloz2001}. 
The simplicity and extensibility of the \wobble method is useful in this regard, as the model could be modified to fit line asymmetries and true Doppler shifts as separate parameters. 
This would help to disentangle signals caused by photospheric features like starspots from the \RV signal due to reflex motion in a planetary system. 
We discuss this prospect further in Section \ref{s:future}.

\section{Telluric Features}
\label{s:tellurics}

\begin{figure*}[h!]
\centering
\includegraphics[width=6.5in]{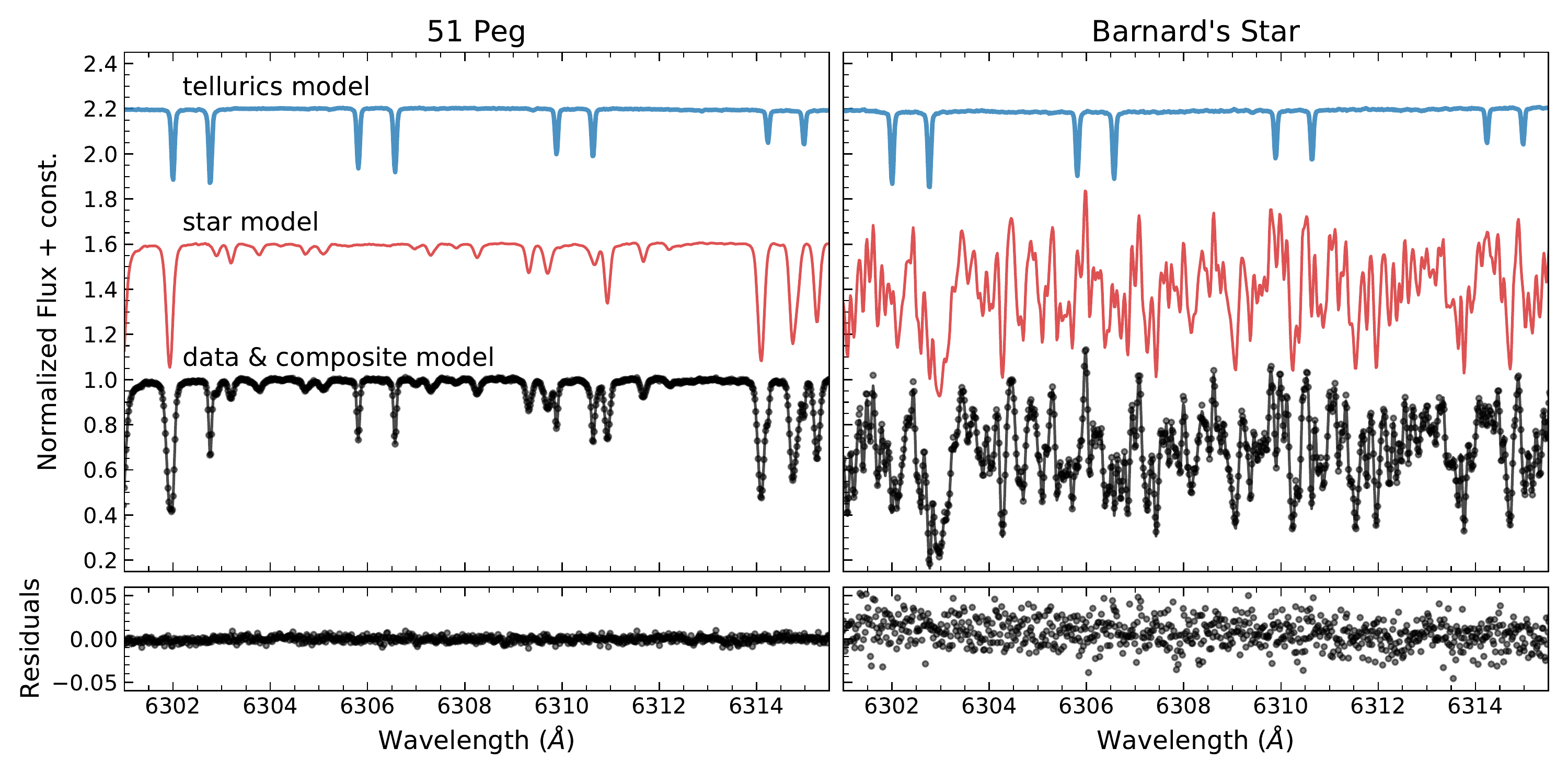}
\caption{Fits to a 15-\ang region with substantial telluric contamination for 51 Peg (left) and \Mdwarf (right). The upper panels show the \wobble best-fit telluric spectrum (blue line), best-fit stellar spectrum (red line), data (black points), and the sum of the telluric and stellar model predictions (black line). Residuals to the fit are shown in the lower panels. Although the fits to each data set were performed entirely independently, the optimized telluric spectra are nearly identical, demonstrating that \wobble successfully finds an accurate representation of the common telluric absorption spectrum.}
\label{fig:telluric_comparison}
\end{figure*}

In the above analyses, telluric spectra were inferred independently for the 51 Peg and \Mdwarf data sets. 
However, in principle the telluric features should be common to all \HARPS spectra, an assumption that we used for the case of HD 189733. 
\changed{As a test of this assumption, we made a comparison between the telluric template spectra derived from the 51 Peg and \Mdwarf data (Figure \ref{fig:telluric_comparison}).} 
Although they were fit using entirely different data sets with dissimilar stellar spectra, the resulting telluric fits are indeed extremely similar. 
This is generally true for all wavelength regions in which tellurics are detected. 
The only regime in which the comparison fails is for low-\SNR spectral orders of the \Mdwarf data, where the telluric spectra are featureless due to the strong regularization, as discussed in Section \ref{s:Mdwarf}. 

The physically motivated expectation that all observations should share a common telluric component could be built into the \wobble model. 
For the fit to HD 189733, for example, we fixed the telluric spectrum to use the model inferred from the better-studied star 51 Peg. 
A more robust, albeit more computationally expensive, approach would be to fit many stars simultaneously so that their shared telluric spectrum can be inferred using all the available data. 
Given enough spectra, this approach should yield an incredibly detailed model of telluric features and their time variability. 
We leave such an effort to future work. 

\begin{figure*}
\centering
\includegraphics[width=5in]{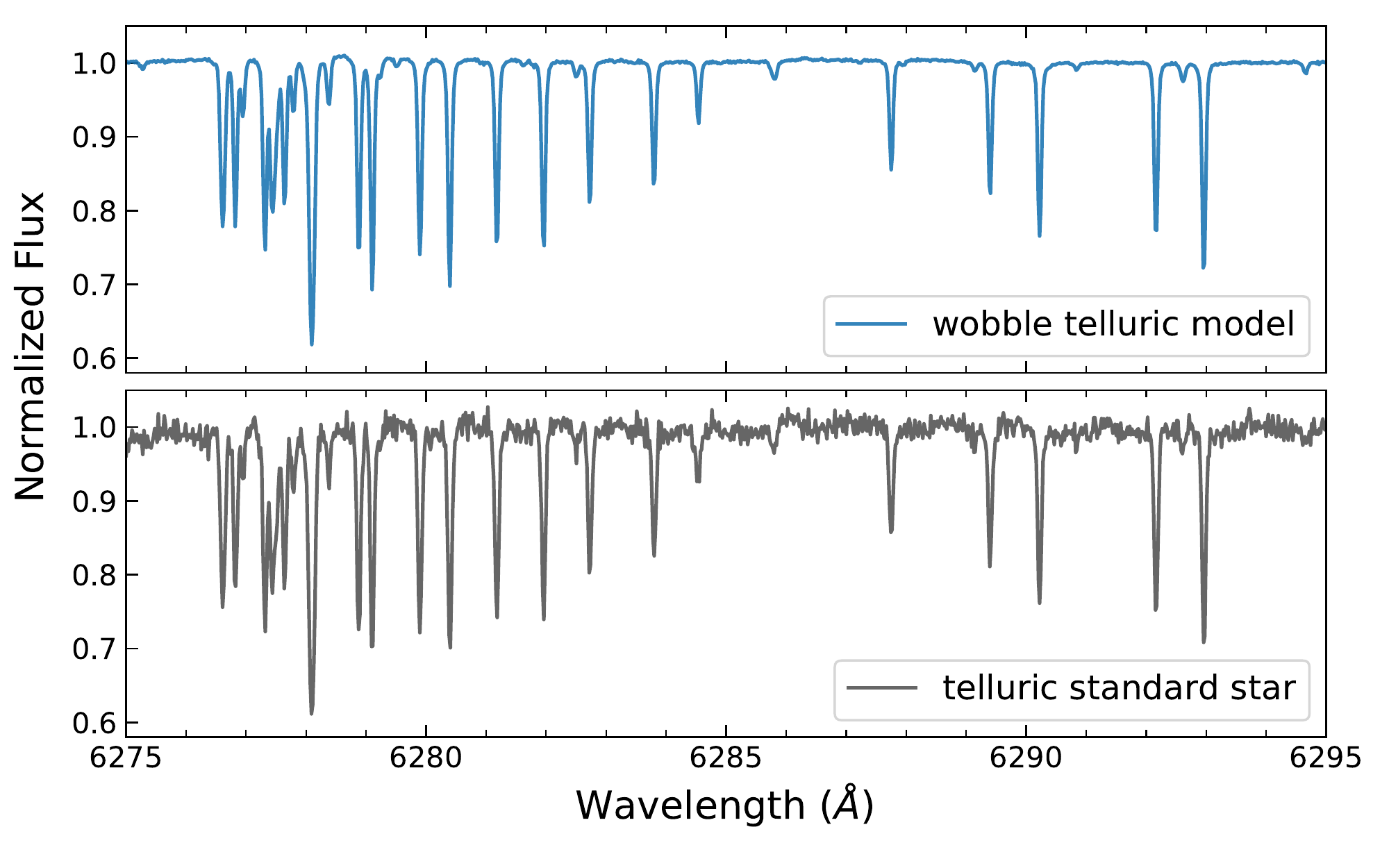}
\caption{Comparison of telluric spectrum derived by \wobble analysis of 51 Peg spectra (upper panel) and the spectrum observed with a telluric standard star (lower panel) around an H$_2$O absorption band. No telluric standard star data was used in determining the \wobble telluric template. The \wobble method delivers a very high \SNR telluric spectrum with no overhead observing time required.}
\label{fig:telluric_standard}
\end{figure*}

As another test of our derived telluric spectra, we compared the mean template from 51 Peg to an observation of the telluric standard star HR 3090 taken by \HARPS at an \SNR of 130. 
As shown in Figure \ref{fig:telluric_standard}, the spectra agree well. 
Moreover, the composite telluric spectrum inferred from the time series of 51 Peg observations is at a much higher \SNR than the single shot telluric standard spectrum, so that low-amplitude lines which border on statistical insignificance in the standard star show up clearly in the \wobble results. 
Unlike a traditional telluric standard observation, no overhead time is required to produce these results.

Of course, one strength of a telluric standard star is that its spectrum can capture the true telluric spectrum at any given moment in time, while the telluric template determined by \wobble is time-averaged and may not perfectly capture the telluric absorption lines in any given observation. 
For this reason, we included time-variable components in the model. 
A physical interpretation of these components should be possible.

\begin{figure*}
\centering
\includegraphics[width=5in]{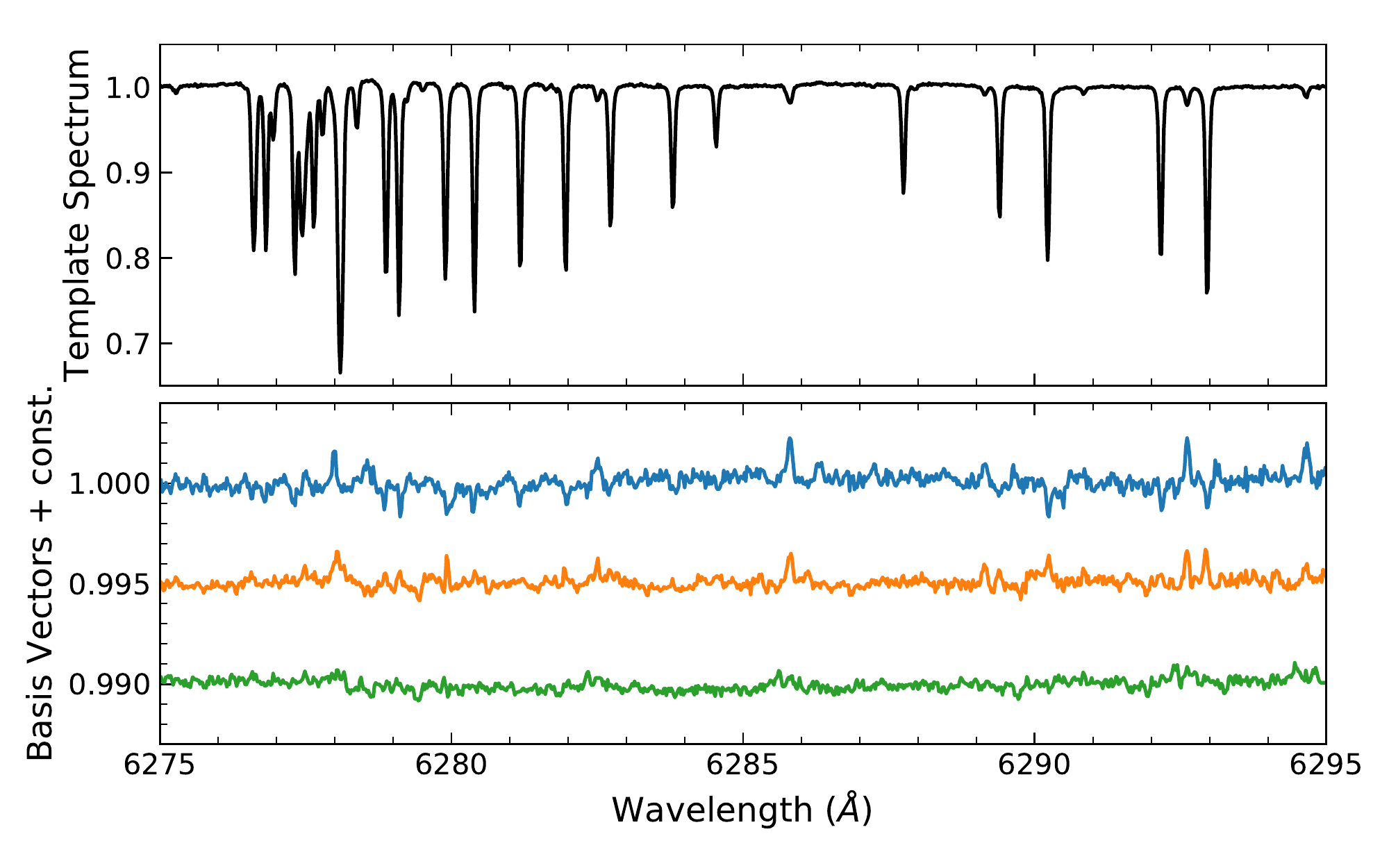}
\caption{Example of the mean telluric spectral template in a small wavelength region (upper panel) and the three basis vectors used to capture the time-variable contributions to the spectrum, staggered by an arbitrary offset for display (lower panel).}
\label{fig:telluric_basis}
\end{figure*}

An example of the inferred variability from the 51 Peg analysis is shown in Figure \ref{fig:telluric_basis}. 
There is clearly coherent structure in the variations. 
As we might expect, many of the lines co-vary with each other: as all known telluric lines in this region arise from H$_2$O, this likely corresponds to time-variable levels of atmospheric water vapor content. 
This hypothesis could be tested in the future by seeking correlations between basis weights and externally-measured weather diagnostics. 

\section{Limitations and Extensibility of \wobble}
\label{s:future}

In this work we have tested \wobble exclusively using \HARPS data, which allowed us to make relatively strict assumptions about the data quality. 
We have also made a number of restrictive assumptions about the underlying physical model, for example taking each astronomical source to be a single star. 
However, there are many other \EPRV data sets which violate some of these assumptions but could nevertheless benefit from the analysis techniques used by \wobble.
 
We now turn our discussion to ways in which other data sets may violate the assumptions made in this work and the corresponding modifications to \wobble that would enable it to operate more effectively in these regimes. 
For simplicity, we break this discussion down into two categories: changes to the quality or type of spectroscopic data used and changes to the model. 
We additionally outline some general changes or enhancements that could be made to potentially improve the \RV precision achieved by \wobble. 

\subsection{Changes to the Data}
\label{s:data-changes}

In general, the data used in this work are high quality in terms of \SNR, spectral resolution, and quantity. 
As demonstrated for the case of \Mdwarf, \wobble is limited in its operability when the data are low in \SNR. 
\wobble implicitly assumes Gaussian noise in logarithmic flux, which is not strictly accurate. 
This could become a substantial issue at low \SNR, making \wobble a poor tool choice for accurate modeling of very low \SNR spectra.
At this point, it is also the case that our continuum-normalization method also does poorly at low \SNR, so there are various reasons \wobble is not optimal for faint sources or very short exposure times.
However, the fact that the model is successful in the form of a linear additive model in the log space means that it could probably be converted to work as a multiplicative model in the linear space; that conversion and investigation is beyond the scope of the current work, which is optimized for performance on typical \HARPS observing campaigns.

Our method may also fail at low spectral resolution. 
In particular, if the stellar spectrum does not shift with respect to the telluric spectrum by at least a resolution element (naively this is expected to become a problem at resolutions below $10^4$), it could become challenging to disentangle the two components.
That said, there still is a causal difference in the data space between signals that are fixed with respect to the star and signals that are fixed with respect to the atmosphere; in principle it is a question not purely of resolution but a combination of resolution and \SNR.
Naively, at low resolution, the model is expected to work well when the total signal-to-noise in the data set (greatly) exceeds the dimensionless ratio of the line width to the radial-velocity variation.

Along with minimum requirements on the spectral data quality, \wobble has some limitations driven by the number of spectra available and the sampling of those spectra. 
Some minimum quantity of spectra are necessarily simply because the spectral templates contribute many free parameters, driving up quantity of data needed for inference. 
The sampling of those spectra are also important because the stellar lines must undergo substantial \RV variations to disentangle them from the telluric features, which for most targets can only be achieved by observing throughout the year to take advantage of the $\sim 30$ \kms~\BERV shift. 
This requirement could be partially mitigated by modeling multiple stars with a shared telluric spectral template, which we discuss in Section \ref{s:improvements}. 
However, in general the \wobble method fundamentally requires a substantial number ($N \gtrsim 10$) of high quality spectra to perform reliably.

Another key assumption made in this work that there is no gas cell represented in the data. 
While this is true for \HARPS, other instruments such as \acronym{HIRES}, \acronym{PFS}, and \acronym{APF} include a gas cell set in optical path to imprint its absorption spectrum on the observed spectra \citep{Butler1996, Crane2010, Vogt2014}. 
Applying \wobble to spectra that include the imprint of a gas cell is technically trivial. 
The gas absorption lines are fixed features at the observatory rest frame which multiply into the spectrum. 
This makes them nearly indistinguishable from the mean telluric spectrum, and they should in fact be absorbed into this component. 
If the observations in question cover a range of different airmasses, it may be necessary to explicitly add a third model component to represent the gas cell; this component would be identical to the tellurics without the airmass scaling. 

Of course, absorption cells in general exist because the instantaneous calibration of the spectrograph in question is not reliable at the required \RV precision level. 
This means that a gas cell instrument will generally violate other, more critical assumptions about the data quality. 

It is likely that for such an instrument, our assumption that the wavelength solution is perfect does not hold; hence the cell. 
One simple way to deal with this would be to allow the observatory rest frame component(s) of the model to change in \RV. 
The stellar \RV measurement would then be differential with respect to the effective spectrograph \RV. 
This effective spectrograph \RV may vary in different spectral orders or smaller wavelength regions.
The most extreme extension would be to include corrections to the wavelength solution as model parameters, and optimize them along with the spectral components and radial velocities. 
This would rely on the use of telluric features as fixed calibration sources \citep{Seifahrt2010}.

Another way a less well-calibrated instrument may violate \wobble's assumptions is in the non-negligible variability of all spectral lines due to changes in the instrumental line spread function (\LSF). 
In principle, if the changes to the \LSF were well-understood they could be hard-coded into a linear operator like the $P$ operator used in Equation \ref{eqn:star}. 
\changed{In this case the \LSF operator should be applied to both the stellar and telluric component models after combining them. Indeed, technically the current \wobble model does not deal correctly with line broadening: the templates we infer for star and tellurics are post-broadening spectra, but this goes against the correct order of operations if line broadening is dominated by instrumental effects that are applied to the \textit{combined} spectra. We have assumed that this effect is negligible for the purposes of this work, but it should certainly be considered more carefully in the future.} 

Realistically speaking, if \LSF variations are present they are probably not sufficiently well-understood to \changed{take a prescriptive approach to their behavior and hard-code them into the form of the $P$ operator}; instead, we might want to fit these variations \changed{using a data-driven approach, which would make the \wobble model significantly non-linear}. 
This leads us into the territory of making more fundamental changes to the \wobble model.

\subsection{Changes to the Model}
\label{s:model-changes}
 
The test data used in this work were all instances of a single, bright star, so that only one stellar component was necessary to model the astronomical source. 
Other data sets may require a more complex model, including additional stars, planetary spectra, or non-negligible sky background. 
They may also require more complex treatment of spectral variability. 
There are some trade-offs associated with adding these components to the model, which we discuss here.

One important feature of \wobble is that it models all fluxes in the log-space. 
When the model consists solely of multiplicative components, as is the case for a single star + telluric absorption, this choice makes the optimization convex at fixed \RV and therefore computationally tractable; it makes possible our iterative optimization scheme.
If we chose instead to convert the $y$ data vectors to linear fluxes and use a model consisting only of additive components, the same equations and algorithms would apply.
Example applications of such a model might be a multiple star system or a faint star with substantial sky background present in the observed spectra. 
If, however, the model needed to incorporate both additive and multiplicative
components as free variables, larger structural changes would be needed in the code and the convexity of the template optimization would be lost. 
Further tests would be needed to determine whether this loss of convexity would slow down the optimization or make finding good optima hard or impossible; it seems likely that it will slow the code but not be impossible.

One extra model component that is at present both simpler to implement and perhaps more important to the results than generic additive components is the spectral continuum. 
We have assumed that our simple continuum normalization is good, but this is certainly not true at a detailed level. 
To some extent, this wrongness is absorbed into the telluric spectrum, as both the tellurics and the continuum can be treated as observatory rest frame features. 
However, since the continuum and tellurics vary through time in different ways, the continuum should in principle be its own independent model component. 
This component will be highly degenerate with the tellurics, but a clever representation could circumvent this: for example, the continuum spectrum could be placed on an extremely sparse wavelength grid to ensure that very localized effects like absorption lines are not included.

Another way in which the \wobble model may fail for some regimes is in its treatment of spectral variability.
We have assumed that the telluric spectrum is variable in a low-dimensional sense only. 
While this assumption has a basis in the physical understanding of telluric lines, which are produced by only a small number of molecular species, it is not guaranteed that this reasoning translates into low variability in the current parameterization. 
We may instead wish to change the spectral representation so that additional physics is included, as discussed further in Section \ref{s:improvements}. 
Alternatively, we could keep the current representation but fit the telluric model to the residuals away from some physics-based model. 

We have also assumed that the stellar spectrum is constant, which is almost certainly false. 
A variety of physical effects cause changes in the spectrum including stellar oscillations, convection, and activity in the form of starspots or plages. 
\wobble can model stellar variability trivially by making the stellar component of the model take a similar form to the tellurics, where a set of basis vectors and weights are inferred from the data to capture variability. 
This \PCA-like approach may be quite effective, since stellar features with similar line formation physics likely co-vary in the spectrum \citep{Davis2017, Dumusque2018}. 
It does, however, introduce potential degeneracies between the inferred stellar variability and the Doppler shift. 
To avoid this degeneracy, we might instead represent the stellar spectrum with an explicit parameterization of the line profile and restrict variability to the line depths, widths, and skews; such a change would fall along the same lines as the global \LSF parameterization proposed in Section \ref{s:data-changes}.

\subsection{Basic Adjustments}
\label{s:improvements}

Even working within \wobble's current assumptions and running on \HARPS-like data, there are potential changes to be made to the implementation of \wobble that may improve its performance. 
Many of these are quite basic and straightforward changes to be made within the existing framework of \wobble. 
We discuss some potential adjustments here.

Perhaps the most obvious shortcoming of \wobble is its reliance on regularization. 
We chose this framework because it keeps the model convex and its implicit assumptions (that the spectra are a flat continuum by default) closely mirror our physical understanding. 
However, setting the regularization amplitudes optimally is by far the computationally slowest and most unwieldy part of applying \wobble to a new data set. 
Dealing with the possibility of overfitting is an inevitable requirement of any model, especially one that relies on as many free parameters as \wobble. 
That being said, in principle as data sets get larger, regularization becomes less crucial. 
One way to lessen the need for telluric regularization might be to run on a data set that is composed of many different stars, each with their own individual spectra and \RVs but with a shared telluric component, so that the data informing the telluric lines is sufficiently large. 
This could be done with relatively trivial changes to the \wobble code, although it would require considerable computational resources.

One choice that \wobble makes differently from other data-driven \RV codes is the way we represent the underlying spectral components. 
We use an extremely simple representation: a model grid of wavelengths and corresponding fluxes that gets linearly interpolated. 
One possible improvement would be to go to a higher-order interpolation scheme. 
Another approach would be a non-parameteric method like a Gaussian process \citep[as in][]{Czekala2017}. 
A third approach would be to use a model that incorporates more physics: this could range from a model with fixed lines that may vary only in terms of common line profile parameters to a complex model including various atomic and molecular physics of the lines. 
A sensible way to implement this within the existing \wobble framework might be to add a physically motivated mean spectral template and optimize $y_{\star, n}$ vectors of residuals away from this template. 
In any case, the general approach of \wobble does not depend intrinsically on the spectral representation, and swapping out the current representation for something else is simple, provided that the representative function can be implemented in \TF.

Another choice we made which is likely suboptimal is the treatment of each spectral order in the echelle data as an independent spectrum with its own stellar \RV. 
In reality, every stellar \RV should be informed by all spectral orders simultaneously. 
Implementing this would take on some extra computational cost, but being in \TF will help to mitigate this. 
However, the reality of color-dependent atmospheric dispersion effects means that in practice each order may have a slightly different effective \RV, since the flux-weighted average observation times will be different \citep{Blackman2017}. 
Thus for some situations allowing the orders to have independent \RV estimates may be the best choice.

While \wobble optimizes a sensible objective function with good properties, it has no direct Bayesian interpretation. 
The regularizations (especially the L1) do not make much sense as prior beliefs, and the output is a point estimate, not a probability or probability density. 
Furthermore, we may at some point want \RV measurements that are the result of \emph{marginalizing out} the spectral and telluric models. 
Right now we don't expect such a full subjective Bayesian treatment to be computationally possible.
However, there certainly has been interesting work in this area, and it might become possible in the near future \citep{Czekala2015}.

Finally, a change that is quite major but worth stating regardless is the possibility of fitting the data directly in the 2D domain. 
Widely used methods of extracting 1D spectra almost certainly sacrifice some degree of spectral information \citep{Bolton2010}.
In the future, we may wish to move to fitting the data directly without extracting. 
The equations for \wobble would change very little in this regime: essentially, the $P$ operator in Equation \ref{eqn:star} would become a 2D interpolator. 
However, implementation of this model would become much more complicated, and extra variable components like sky background and spectrograph dispersion functions would need to be introduced. 
While this implementation is far beyond the scope of current work, we nevertheless emphasize that the \wobble model is a useful starting framework in which to work towards this goal.

\section{Conclusion}
\label{s:conclusion}

Extremely precise \RV measurements are a critical part of modern exoplanet detection and characterization. 
Many dedicated \EPRV instruments exist and more are being built with the goal of achieving better than \ms precision over long timescales. 
Beyond the hardware challenges of building such instruments, software challenges exist as well: extracting maximally precise \RVs from high-quality spectra is not a trivial task. 
Moreover, most existing \EPRV pipelines are closed-source, proprietary, and built for use with a specific instrument in mind. 
While traditional \RV extraction methods are demonstrably suboptimal when it comes to treatment of telluric features and use of a fixed stellar template spectrum, experimenting with changes to the existing pipelines is generally not possible. 

\changed{The data from these \EPRV campaigns are also useful for the purpose of precise stellar characterization using the co-added spectra. 
Since such a co-added spectrum is not a primary intended data product of \EPRV instrument pipelines, this is generally left to the user, who is again limited by the closed-source nature of most pipelines in their ability to take properly into account the true uncertainties in producing a net spectrum, or in removing the telluric absorption features that introduce noise to the stellar spectrum.}

In this work, we have proposed a simple linear model for simultaneously inferring stellar spectra, telluric spectra, and \RVs from the data. 
Our model does not rely on physical knowledge of the star or the Earth's atmosphere. 
\changed{It is designed with the dual aims of producing precise \RV estimates and robust stellar spectral templates from the data.}
We implement this method in \wobble, an open-source code designed to be extensible and adaptable for a variety of data sets. 

By running the \wobble algorithm on archival \HARPS spectra for a variety of stars, we have demonstrated its basic capabilities. 
The \RVs achieved through this method are comparably precise to those obtained with the closed-source \HARPS pipeline. 
Furthermore, the stellar and telluric spectra inferred purely from the data are high-quality and accurate. 
We have shown that consistent telluric features are found from independent data sets with extremely different properties. 

The data-driven methods used in \wobble are highly promising for the treatment of microtelluric lines and other deviations from spectral models. 
This makes the approach particularly valuable for upcoming infrared surveys targeting M dwarfs, where telluric features are non-negligible and stellar models often fall short of matching the data. 
It also eliminates costs in observational overhead for obtaining telluric standard spectra or accurate stellar templates.

We make \wobble freely available to the \RV community to be adapted for use with various instruments and data sets. 
We highlight the versatility of the algorithm in Section \ref{s:future} with suggested changes and improvements for applications to different regimes. 
It is our hope that openly shared development of next-generation \changed{spectral} analysis techniques will enable the community as a whole to make full use of the information-rich data being produced by current and future \EPRV surveys.

\software{%
    astropy \citep{Astropy13, Astropy18},
    numpy \citep{Van-Der-Walt:2011},
    matplotlib \citep{Hunter:2007},
    scipy \citep{Jones:2001},
    tensorflow \citep{Abadi15}, 
    exoplanet \citep{exoplanet},
    starry \citep{Luger2018},
    pymc3 \citep{pymc3},
    theano \citep{theano}
}

\acknowledgements{\changed{We gratefully acknowledge the referee, Ian Czekala, for a thoughtful review that substantially improved this work.}
It is a pleasure to acknowledge Jacob L. Bean, John M. Brewer, Heather Knutson, Timothy Morton, Melissa Ness, Andreas Seifahrt, Julian St\"{u}rmer, \changed{and Sharon Xuesong Wang} for helpful discussions. 
We are grateful to Hans-Walter Rix and the Max-Planck-Institut f\"ur Astronomie for their hospitality while developing key parts of this project. Special thanks go to Geert Barentsen, Michael Gully-Santiago, Christina Hedges, Susan Mullaly, and Z\'{e} Vin\'{i}cius for their vital contribution to the most important part of \wobble's documentation (the pronunciation guide).

This research has made use of the services of the ESO Science Archive Facility. 

This work has made use of data from the European Space Agency (ESA) mission
{\it Gaia} (\url{https://www.cosmos.esa.int/gaia}), processed by the {\it Gaia}
Data Processing and Analysis Consortium (DPAC,
\url{https://www.cosmos.esa.int/web/gaia/dpac/consortium}). Funding for the DPAC
has been provided by national institutions, in particular the institutions
participating in the {\it Gaia} Multilateral Agreement.

Work by B.T.M. was performed under contract with the Jet
Propulsion Laboratory (JPL) funded by NASA through
the Sagan Fellowship Program executed by the NASA
Exoplanet Science Institute.}

\bibliographystyle{apj}
\bibliography{paper.bib}

\end{document}